\documentclass[RNAAS]{aastex62}

\newcommand{\Msun}{{\rm M_{\odot}}}

\newcommand{\mpc}{\, {\rm Mpc}}
\newcommand{\kpc}{\, {\rm kpc}}

\newcommand{\kmps}{\, {\rm km \, s^{-1}}}

\newcommand{\Myr}{\,{\rm Myr}}

\newcommand{\Omegap}{\,\Omega_{\rm p}}

\shorttitle{Long GMC Lifetimes}
\shortauthors{Koda}

\begin{document}
\setcounter{table}{0}

\title{Long GMC Lifetimes: Using the Method and Data of Meidt et al. 2015 with A Correction}

\correspondingauthor{Jin Koda}
\email{jin.koda@stonybrook.edu}

\author{Jin Koda}
\affil{Department of Physics and Astronomy, Stony Brook University, Stony Brook, NY 11794-3800}

\begin{abstract}
\citet{Meidt:2015vy} derived short lifetimes of 20-30~Myr for giant molecular clouds (GMCs) in M51.
Their novel approach utilizes a decline of the GMC population during their inter-arm passage
from one spiral arm to the next.
Using the inter-arm travel time $t_{\rm travel}$ as a fiducial clock,
they converted the decline rate to GMC lifetimes.
They implicitly adopted zero as the pattern speed of the spiral arms,
resulting in a very short $t_{\rm travel}$.
However, it is well established that the spiral arm pattern in M51
is rotating \citep{Meidt:2008vt, Meidt:2013to},
and that $t_{\rm travel}$ should be measured with respect to the rotating spiral pattern.
Here we use the same method and data of \citet{Meidt:2015vy}
and re-evaluate GMC lifetimes by accounting for the pattern speed given by \citet{Meidt:2013to}.
This correction gives a longer $t_{\rm travel}$,
and consequently longer GMC lifetimes of 60-500~Myr.
\end{abstract}


\section{Introduction} \label{sec:intro}
\citet{Meidt:2015vy} derived GMC lifetimes $t_{\rm GMC}$ in M51 as a fraction of their inter-arm travel time $t_{\rm travel}$.
They estimated $t_{\rm travel}$ as half the disk rotation timescale for the two-armed spiral galaxy,
\begin{equation}
    t_{\rm travel} = \frac{\pi R}{V_{\rm rot}}, \label{eq:ttravel1}
\end{equation}
where $R$ is the galactocentric radius and $V_{\rm rot}$ is the rotation velocity of the disk.
However, it has been established that the spiral pattern in M51 is rotating
\citep[e.g., ][]{Meidt:2008vt, Meidt:2013to}.
$t_{\rm travel}$ should be calculated with respect to the spiral arms
in the rotating frame of the spiral pattern as,
\begin{equation}
    t_{\rm travel} = \frac{\pi R}{V_{\rm rot}-\Omegap R}, \label{eq:ttravel2}
\end{equation}
where $\Omegap$ is the pattern speed.
We re-evaluate GMC lifetimes with the method and data of \citet{Meidt:2015vy}, but with
this correction on $t_{\rm travel}$.

\section{Parameters} \label{sec:param}

We adopt all the notations and parameters of \citet{Meidt:2015vy}
except for $\Omegap$ which is from \citet{Meidt:2013to}.
They adopted a distance to M51 of $D=7.6\mpc$.
All the parameters are listed in Table \ref{tab:lifetime}.

\citet{Meidt:2015vy} performed their analysis in the radius range of $R\sim 40$-$80\arcsec$ (1.5-3.0$\kpc$).
\citet{Meidt:2008vt} suggested a radial decline of the pattern speed $\Omegap$ in M51;
as updates in \citet{Meidt:2013to}, 
$\Omegap$ is high at $90\kmps\,\kpc^{-1}$ out to its corotation radius of $R_{\rm CR}=55\arcsec$ ($2.0\kpc$)
and transitions to $\Omegap=55\kmps\,\kpc^{-1}$ outside.
\citet{Meidt:2013to} explained this transition as due to an overlap of the corotation of the inner spiral
pattern and the inner 4:1 resonance of the outer pattern.

A constant rotation velocity $V_{\rm rot}$ is assumed based on the rotation curve in this radius range
\citep[their Figure 10]{Meidt:2013to}.
From $\Omegap=90\kmps\,\kpc^{-1}$ and its corotation radius $R_{\rm CR}=55\arcsec$,
we derive $V_{\rm rot}=182\kmps$.
The second corotation radius corresponding to $\Omegap=55\kmps\,\kpc^{-1}$
is $R_{\rm CR}=3.3\kpc$, which is outside the radius range of their analysis.

The inter-arm travel velocity with respect to the spiral pattern is $V_{\rm rot}-\Omegap R$.
The inter-arm travel time $t_{\rm travel}$ is calculated with eq. (\ref{eq:ttravel2}).

\citet{Meidt:2015vy} measured the fraction of lost GMC population $F_{\rm lost}$,
i.e., the decline rate of GMC population during inter-arm passage.
We read the values of $F_{\rm lost}$ at different radii from Figure 7 (right) of \citet{Meidt:2015vy}.

\citet{Meidt:2015vy} calculated the GMC population lifetime $\tau$ as
\begin{equation}
    \tau = \frac{t_{\rm travel}}{2} \frac{1}{F_{\rm lost}}. \label{eq:life}
\end{equation}
This includes the decline and growth of the population,
\begin{equation}
    \frac{1}{\tau} = \frac{1}{\tau_{\rm true}} - \frac{1}{\tau_{\rm grow}}, \label{eq:tau}
\end{equation}
where $\tau_{\rm true}$ is the time it would take to reduce the initial population,
and $\tau_{\rm grow}$ represents the time to increase the population.
\citet{Meidt:2015vy} assumed $\tau_{\rm grow} \gg \tau_{\rm true}$ and adopted $\tau$
as the GMC lifetimes.
We follow this approach as it is unlikely that a GMC population grows significantly
in inter-arm regions.

\section{Results} \label{sec:results}

Table \ref{tab:lifetime} list the results.
The travel time of \citet{Meidt:2015vy} is estimated to be $t_{\rm travel}=34 (R/{\rm 2 kpc}){\rm Myr}$
using eq. (\ref{eq:ttravel1}) and $V_{\rm rot}=182\kmps$
-- this calculation is roughly consistent with the $t_{\rm travel}$ line in Figure 7 of \citet{Meidt:2015vy}.
The corrected $t_{\rm travel}$ in Table \ref{tab:lifetime} (with eq. \ref{eq:ttravel2})
are much longer, by a factor of 3-16, than their $t_{\rm travel}$.
The long travel time suggests that
if the GMC lifetimes were as short as the originally reported (20-30 Myr),
they would not be able to move much across the inter-arm regions.
The correction with $\Omegap$ gives long GMC lifetimes of $\tau = 60$-$500\mpc$, much longer than those originally reported.

We note that the same method and data of \citet{Meidt:2015vy} are used here
with one correction on the pattern speed.
The same limitations also apply here.
For example, the lifetimes calculated here are for the GMCs detected
above their sensitivity limit ($>10^5\Msun$).
This massive GMC population accounts for only about half of the CO emission in M51,
and the other half arises from GMCs smaller than the detection limit.
This method is vulnerable to variations of GMC properties during inter-arm passages.
For example, if they become fainter (colder) during the passages,
a large population may sink below the detection limit,
resulting in an overestimation of $F_{\rm lost}$ and underestimation of $\tau$.

\floattable
\begin{deluxetable}{ccccccc}
\tablecaption{GMC Lifetimes ($\tau$) and Other Parameters\label{tab:lifetime}}
\tablehead{
\multicolumn{2}{c}{R} & \colhead{$F_{\rm lost}$} & \colhead{$\Omegap$} & \colhead{$V_{\rm rot}-\Omegap R$} & \colhead{$t_{\rm travel}$}  & \colhead{$\tau$}  \\
\cline{1-2}
\colhead{(arcsec)} & \colhead{(kpc)} & \nocolhead{} & \colhead{($\kmps \kpc^{-1}$)}   & \colhead{($\kmps$)} & \colhead{($\Myr$)} & \colhead{($\Myr$)}
}
\startdata
44.4	&	1.64	&	0.75	& 90 &	35	&	143		&	95		\\
51.6	&	1.90	&	0.52	& 90 &	11	&	517		&	497		\\
58.8	&	2.17	&	0.82	& 55 &	63	&	105		&	64	    \\
66.0    &	2.43	&	0.76	& 55 &	49	&	153		&	101		\\
73.2	&	2.70	&	0.40	& 55 &	34	&	243		&	304		\\
80.4	&	2.96	&	0.67	& 55 &	19	&	467		&	348		\\
\enddata
\tablecomments{
Following \citet{Meidt:2013to} (and \citet{Meidt:2008vt}),
we adopt $\Omegap=90\kmps\,\kpc^{-1}$ in $R<2.0\kpc$ and $55\kmps\,\kpc^{-1}$ outside.
We assume a constant rotation velocity of $V_{\rm rot}=182\kmps$,
which is calculated from the $\Omegap$ and $R_{\rm CR}$ given by \citet{Meidt:2013to}.
}
\end{deluxetable}

 \acknowledgments
JK acknowledges support from NSF through grants AST-1812847 and AST-2006600.


\begin{thebibliography}{}
\expandafter\ifx\csname natexlab\endcsname\relax\def\natexlab#1{#1}\fi

\bibitem[{{Meidt} {et~al.}(2008){Meidt}, {Rand}, {Merrifield}, {Shetty}, \&
  {Vogel}}]{Meidt:2008vt}
{Meidt}, S.~E., {Rand}, R.~J., {Merrifield}, M.~R., {Shetty}, R., \& {Vogel},
  S.~N. 2008, \apj, 688, 224

\bibitem[{{Meidt} {et~al.}(2013){Meidt}, {Schinnerer}, {Garc{\'\i}a-Burillo},
  {Hughes}, {Colombo}, {Pety}, {Dobbs}, {Schuster}, {Kramer}, {Leroy}, {Dumas},
  \& {Thompson}}]{Meidt:2013to}
{Meidt}, S.~E., {Schinnerer}, E., {Garc{\'\i}a-Burillo}, S., {et~al.} 2013,
  \apj, 779, 45

\bibitem[{{Meidt} {et~al.}(2015){Meidt}, {Hughes}, {Dobbs}, {Pety}, {Thompson},
  {Garc{\'\i}a-Burillo}, {Leroy}, {Schinnerer}, {Colombo}, {Querejeta},
  {Kramer}, {Schuster}, \& {Dumas}}]{Meidt:2015vy}
{Meidt}, S.~E., {Hughes}, A., {Dobbs}, C.~L., {et~al.} 2015, \apj, 806, 72

\end{thebibliography}
\end{document}